\documentclass{article}
\title{An Explanation of \emph{Spin} Based on Classical Mechanics and Electrodynamics}
\author{O. Chavoya-Aceves\\
Camelback H. S., Phoenix, Arizona, USA.\\
chavoyao@yahoo.com}

\begin{document}
\maketitle

\begin{abstract}
It is proved that, according to Classical Mechanics and
Electrodynamics, the trajectory of the center of mass of a neutral
system of electrical charges can be deflected by an inhomogeneous
magnetic field, even if its internal angular momentum is zero.
This challenges the common view about the function of the
Stern-Gerlach apparatus, as resolving the eigen-states of an
intrinsic angular momentum. Doubts are cast also on the supposed
failure of Schr\"odinger's theory to explain the properties of
atoms in presence of magnetic fields without introducing
\emph{spin} variables.

\end{abstract}
\section{Introduction}
In a previous paper \cite{OCHAVOYA} we have shown that the
correspondence principle implies that the operator for angular
momentum of a particle in an electromagnetic field is:
\begin{equation}\label{operadores de momento angular}
  \hat{L}=\vec{r}\times(-i\hbar \nabla -\frac{q}{c}\vec{A}),
\end{equation}
where $\vec{A}$ is the vector potential. As we mentioned there,
(\ref{operadores de momento angular}) is also required to
guarantee that the corresponding expected values are
gauge-invariant.

Under a gauge transformation that transforms the electrodynamic
potentials in the form:
\begin{equation}\label{transformacion de norma}
  \phi'= \phi +\frac{1}{c}\frac{\partial \Lambda}{\partial t}
  \ \ \ \ \vec{A}'= \vec{A}+\nabla \Lambda
\end{equation}
the wave function \cite{WEYL} is transformed as
\begin{equation}\label{transformacion de la funcion de onda}
 \psi'=e^{-\frac{iq}{\hbar c}\Lambda}\psi.
\end{equation}

The expected value of the operator $-i\hbar \vec{r}\times\nabla$
for the wave function $\psi'$ is
\begin{equation}\label{valor esperado del momento angular}
  \int \psi'^\star(-i\hbar\vec{r}\times\nabla)\psi'=%
   \int \psi^\star(-i\hbar\vec{r}\times\nabla)\psi+\frac{q}{c}\int
   \psi^\star\psi\vec{r}\times\nabla\Lambda,
\end{equation}
that is not independent of $\Lambda$, showing that the operator
$-i\hbar\vec{r}\times\nabla$ cannot represent a physical
observable, but where there is not a magnetic field.

Also, from the general relation
\begin{equation}\label{operador derivada}
  \frac{d{\hat{f}}}{dt}=\frac{i}{\hbar}[\hat{H},\hat{f}]+\frac{\partial \hat{f}}{\partial
  t},
\end{equation}
it's clear that the term $-\frac{q}{c}\vec{r}\times\vec{A}$ must
be included as part of the angular momentum, if the correct
contribution of the electric field to the torque is going to be
obtained.

Thus, we are forced to conclude that (\ref{operadores de momento
angular}) is the correct form of the angular momentum and,
consequently, that the eigenvalues and eigenfunctions of angular
momentum depend of the configuration of electromagnetic field.
(For example: the eigenvalues of angular momentum are not integral
multiples of $\hbar$, but where there is not a magnetic field.)
This calls for a revision of the theory of angular momentum, the
theory of spin, in particular, and the theory of interaction of
atoms with magnetic fields, that we undertake here, for the
hydrogen atom, from the classical models to the corresponding
quantum equations.

The first thing we'll note is that, following a classical
lagrangian approach, it can be proved that the motion of the
center of mass and the internal motion of a neutral system of
electrical charges are not physically independent in presence of a
magnetic field. From this we'll prove that the classical
trajectory of the center of mass of a neutral system of electrical
charges can be deflected by an inhomogeneous magnetic field, even
if its internal angular momentum is zero. This deflection is also
predicted by Schr\"odinger theory, in view of Ehrenfest's theorem,
challenging the common belief about the function of the
Stern-Gerlach apparatus, as resolving the eigen-states of an
intrinsic angular momentum.

Also, we'll see that the main evidence we have of the failure of
Schr\"odinger's theory to explain the properties of atoms in
presence of magnetic fields is not completely reliable, because
the usual formulation of the problem \cite[p. 541]{MESSIA} is not
accurate. It's based on four assumptions \cite[p. 541]{MESSIA}
\cite[pp. 359-60]{BOHM}:
\begin{enumerate}
  \item That
the operator $-i\hbar\vec{r}\times\nabla$ corresponds to the
angular momentum---in presence of the magnetic field---and,
therefore, that the allowed values of the projection of the
angular momentum along the magnetic field are integral multiples
of $\hbar$.
  \item That the energy of the motion of the center of mass and the
internal energy of a neutral system of electrical charges are
physically independent, even in presence of a magnetic field.
  \item That the energy of interaction with the magnetic field
  can be written in the form:
  \[
  E_{\vec{H}}=-\vec{\mu}\cdot\vec{H}.
  \]
where
\begin{equation}\label{relacion entre momento angular y momento magnetico}
  \vec{\mu}=\frac{e}{2mc}\vec{L}
\end{equation}
\item That the projection of the operator
$-i\hbar\vec{r}\times\nabla$ along the magnetic field represents a
constant of motion.
\end{enumerate}

We have already seen why the first assumption is not sound and
we'll prove, in a very simple, but rigorous way, that the other
three are not true either, for the hydrogen atom.

In the last section we make an analysis of the classical magnetic
field associated to the hydrogen atom, showing that (\ref{relacion
entre momento angular y momento magnetico}) is the result of a
time-average of dynamical variables, which is not suitable for
quantization. Furthermore, at this statistical level, we have a
correction to the gyromagnetic ratio of the internal angular
momentum, since we prove that:
\begin{equation}\label{relacion entre momento angular y momento magnetico dos}
  \vec{\mu}=\frac{e}{2mc}\frac{m_p-m_e}{M}\vec{L}.
\end{equation}

The ideas exposed in this paper support an explanation of the
phenomena associated to \emph{spin} as consequences of the Laws of
Electrodynamics (Lorentz' and Ampere's) as applied to systems of
electrical charges as wholes, but not as manifestations of
intrinsic properties of punctual particles, as was also sustained
in a different way by Bohr, who believed that the \emph{spin} was
only an abstraction, useful to compute the angular
momentum\cite{GARRAWAY-STENHOLM}.
\section{Hydrogen Atom in an Uniform Magnetic Field}
Let's consider the classical lagrangian of a hydrogen atom under
the action of an external uniform magnetic field. The vector
potential can be chosen as:
\begin{equation}\label{vector potential}
  \vec{A}(\vec{r})=\frac{1}{2}\vec{H}\times \vec{r},
\end{equation}
and the Lagrange's Function can be written as:
\begin{equation}\label{primera lagrangiana}
  L(\vec{r}_p,\vec{r}_e,\vec{v}_p,\vec{v}_e)=%
  \frac{1}{2}m_p v_p^2+\frac{1}{2}m_e v_e^2+%
  \frac{e^2}{\|\vec{r}_p-\vec{r}_e\|}+%
  \frac{e}{2c} \vec{H}\cdot((\vec{r}_p\times\vec{v}_p)-(\vec{r}_e\times\vec{v}_e))
\end{equation}

Let's do the substitution:
\begin{equation}\label{introduccion del centro de masas}
  \vec{R}=\frac{m_p\vec{r}_p+m_e\vec{r}_e}{M};\ \ \
  \vec{r}=\vec{r}_e-\vec{r}_p
\end{equation}
(where $M=m_p+m_e$), in such way that:
\begin{equation}\label{sustitucion de la coordenada del proton}
  \vec{r}_p=\vec{R}-\frac{m_e}{M}\vec{r};%
\ \ \ \vec{r}_e=\vec{R}+\frac{m_p}{M}\vec{r}.
\end{equation}

Then we have
\[
\vec{r}_p\times\vec{v}_p=%
\vec{R}\times\dot{\vec{R}}-%
\frac{m_e}{M}\vec{r}\times\dot{\vec{R}}-%
\frac{m_e}{M}\vec{R}\times\dot{\vec{r}}+%
\frac{m_e^2}{M^2}\vec{r}\times\dot{\vec{r}},
\]
and
\[
\vec{r}_e\times\vec{v}_e=%
\vec{R}\times\dot{\vec{R}}+%
\frac{m_p}{M}\vec{r}\times\dot{\vec{R}}+%
\frac{m_p}{M}\vec{R}\times\dot{\vec{r}}+%
\frac{m_p^2}{M^2}\vec{r}\times\dot{\vec{r}}.
\]
Therefore
\[
\vec{r}_p\times\vec{v}_p-\vec{r}_e\times\vec{v}_e=%
-\vec{r}\times\dot{\vec{R}}-\vec{R}\times\dot{\vec{r}}-\frac{m_p-m_e}{M}\vec{r}\times\dot{\vec{r}}
\]
and
\begin{equation}\label{segunda lagrangiana}
  L(\vec{R},\vec{r},\dot{\vec{R}},\dot{\vec{r}})=%
  \frac{1}{2} M \dot{\vec{R}}^2+\frac{1}{2}\mu\dot{\vec{r}}^2+%
  \frac{e^2}{r}+
\end{equation}
\[
\frac{e}{2c}\vec{H}\cdot\left[-\vec{r}\times\dot{\vec{R}}
  -\vec{R}\times\dot{\vec{r}}-\frac{m_p-m_e}{M}\vec{r}\times\dot{\vec{r}}\right],
\]
where $\mu$ is the reduced mass.

The term $-\vec{R}\times\dot{\vec{r}}$ depends of the position of
the center of mass, which is physically unacceptable. However,
given that
\[
-\vec{R}\times\dot{\vec{r}}=-\frac{d(\vec{R}\times\vec{r})}{dt}-\vec{r}\times\dot{\vec{R}},
\]
the function (\ref{segunda lagrangiana}) can be replaced by:
\begin{equation}\label{cuarta lagrangiana}
  L(\vec{R},\vec{r},\dot{\vec{R}},\dot{\vec{r}})=%
  \frac{1}{2} M \dot{\vec{R}}^2+\frac{1}{2}\mu\dot{\vec{r}}^2+%
  \frac{e^2}{r}-
\end{equation}
\[
\frac{e}{c}\vec{H}\cdot(\vec{r}\times\dot{\vec{R}})
-\frac{e}{2c}\frac{m_p-m_e}{M}\vec{H}\cdot(\vec{r}\times\dot{\vec{r}}),
\]
or
\begin{equation}\label{spuky lagrangiana}
  L(\vec{R},\vec{r},\dot{\vec{R}},\dot{\vec{r}})=%
  \frac{1}{2} M \dot{\vec{R}}^2+\frac{1}{2}\mu\dot{\vec{r}}^2+%
  \frac{e^2}{r}-\frac{e}{2\mu c}\vec{H}\cdot(K_L\vec{L}+2 \vec{S}),
\end{equation}
where
\[
K_L=\frac{m_p-m_e}{M}
\]
\[
\vec{L}=\mu\vec{r}\times\dot{\vec{r}},
\]
and
\[
\vec{S}=\mu \vec{r}\times\dot{\vec{R}}.
\]

If not were by the term
\[
\frac{1}{2}M\dot{\vec{R}}^2
\]
(\ref{spuky lagrangiana}) looks like the Lagrange's Function of a
system with an intrinsic angular momentum $\vec{S}$.

The corresponding momenta are:
\begin{equation}\label{momenta}
  P_{\vec{R}}=M\dot{\vec{R}}-\frac{e}{c}\vec{H}\times\vec{r};\ \ \ \vec{p}_{\vec{r}}=%
  \mu \dot{\vec{r}}-%
  \frac{e}{2c}K_L\vec{H}\times\vec{r}
\end{equation}

From this we get the energy, that is a constant of motion:
\begin{equation}\label{energy}
  E=\frac{1}{2}M\dot{\vec{R}}^2+\frac{1}{2}\mu\dot{\vec{r}}^2-\frac{e^2}{r}
\end{equation}

The equations of motion are:
\begin{equation}\label{ecuacion para el centro de masas}
  M\ddot{\vec{R}}=\frac{e}{c}\vec{H}\times\dot{\vec{r}}
\end{equation}
and
\begin{equation}\label{ecuacion para el movimiento relativo}
  \mu
    \ddot{\vec{r}}=-\frac{e^2\vec{r}}{r^3}-%
    \frac{e}{c}\dot{\vec{\rho}}\times\vec{H}
\end{equation}
where
\begin{equation}\label{expresion que aparece en la ecuacion del movimiento}
    \dot{\vec{\rho}}=\dot{\vec{R}}+K_L\dot{\vec{r}}
\end{equation}

Considering equation (\ref{ecuacion para el centro de masas}), we
can realize that the kinetic energy of the center of mass is not a
constant of motion.  Therefore, since the total energy is
conserved, the internal motion and the motion of the center of
mass are not independent: they are coupled nothing less than by
the rules of transformation of electromagnetic fields. Of course,
all of this is classical, but still holds for quantum mechanics.

The Hamilton's Function is:
\begin{equation}\label{hamiltoniana}
  H(\vec{R},\vec{r},\vec{P}_{\vec{R}},\vec{P}_{\vec{r}})=%
   \frac{(\vec{P}_{\vec{R}}+\frac{e}{c}\vec{H}\times\vec{r})^2}{2M}+\frac{(
\vec{p}_{\vec{r}}+%
  \frac{e}{2c}K_L\vec{H}\times\vec{r}
   )^2}{2\mu}-\frac{e^2}{r},
\end{equation}
and the Hamiltonian Operator:
\begin{equation}\label{hamiltoniano}
  \hat{H}=%
   \frac{(-i\hbar\nabla_{\vec{R}}+\frac{e}{c}\vec{H}\times\vec{r})^2}{2M}+\frac{(
-i\hbar\nabla_{\vec{r}}+%
  \frac{e}{2c}K_L\vec{H}\times\vec{r}
   )^2}{2\mu}-\frac{e^2}{r},
\end{equation}

After some algebra and the usual neglection of second order terms,
(\ref{hamiltoniano}) is transformed into:

\begin{equation}\label{hamiltoniano aproximado}
  \hat{H}=-\frac{\hbar^2}{2M}\nabla_{\vec{R}}^2-\frac{\hbar^2}{2\mu}%
  \nabla_{\vec{r}}^2-\frac{e^2}{r}+\frac{e\vec{H}}{2\mu}\cdot(K_L\hat{l}+2\hat{s})
\end{equation}
where
\begin{equation}\label{momento angular}
  \hat{l}=-i\hbar\vec{r}\times\nabla_{\vec{r}},
\end{equation}
and
\begin{equation}\label{spin}
  \hat{s}=-i\hbar\frac{\mu}{M}\vec{r}\times\nabla_{\vec{R}}.
\end{equation}
showing that the usual formulation of the problem---that affords
the main evidence we have of the failure of Schr\"odinger's theory
to explain the properties of atoms in presence of magnetic fields
without introducing \emph{spin} variables---is not completely
reliable, since it is grounded on four assumptions:
\begin{enumerate}
  \item That
the operator $-i\hbar\vec{r}\times\nabla$ corresponds to the
angular momentum---in presence of the magnetic field---and,
therefore, that the allowed values of the projection of the
angular momentum along the magnetic field are integral multiples
of $\hbar$.
  \item That the energy of the motion of the center of mass and the
internal energy of a neutral system of electrical charges are
physically independent, even in presence of magnetic field.
  \item That the energy of interaction with the magnetic field
  can be written in the form:
  \[
  E_{\vec{H}}=-\vec{\mu}\cdot\vec{H}.
  \]
where
\[
  \vec{\mu}=\frac{e}{2\mu c}\vec{L}
\]
\item That the projection of the operator
$-i\hbar\vec{r}\times\nabla$ along the magnetic field represents a
constant of motion.
\end{enumerate}
We have already shown that the first is not sound, in the
introduction. The second and the third are not true, for the
hydrogen atom, as follows from eq. (\ref{hamiltoniano
aproximado}). Finally, We notice that:
\begin{equation}\label{conmutadores}
  [\hat{s}_i,\hat{l}_j]=-i\hbar\frac{\mu}{M}\epsilon_{iab}[r_a,\hat{l}_j]\frac{\partial}{\partial
  R_b}=\hbar^2\frac{\mu}{M}\epsilon_{abi}\epsilon_{ajc}r_c\frac{\partial}{\partial
  R_b}=
\end{equation}
\[
\hbar^2\frac{\mu}{M}(\delta_{bj}\delta_{ic}-\delta_{bc}\delta_{ij})r_c\frac{\partial}{\partial
  R_b}=\hbar^2\frac{\mu}{M}\left(r_i\frac{\partial }{\partial
  R_j}-\delta_{ij}\vec{r}\cdot\nabla_{\vec{R}}\right).
\]
In particular
\[
[\hat{s}_z,\hat{l}_z]=-\hbar^2\frac{\mu}{M}\left(r_x\frac{\partial}{\partial
R_x }+r_y\frac{\partial}{\partial R_y }\right)
\]
Consequently,
\[
[\hat{H},\vec{H}\cdot\hat{l_z}]\neq 0,
\]
showing that the fourth assumption isn't true either.

This and the striking structure of function (\ref{spuky
lagrangiana}) cast serious doubts on the very existence of
\emph{spins} as intrinsic angular momenta.

Actually, as we'll prove in next section, (\ref{spuky
lagrangiana}) is the correct Lagrange's Function for an atom in an
inhomogeneous magnetic field, where $\vec{H}$ is simply replaced
by $\vec{H}(\vec{R})$, and the corresponding equation of motion
for the center of mass is:

\begin{equation}\label{ecuacion del movimiento stern-gerlach}
  M\ddot{\vec{R}}=\frac{e}{c}\vec{H}\times\dot{\vec{r}}+%
\frac{e}{c}[(\dot{\vec{R}}\cdot\nabla_{\vec{R}})\vec{H}]\times\vec{r}-\frac{e}{2\mu
c}\nabla_{\vec{R}} [\vec{H}\cdot(K_L\vec{L}+2\vec{S})]
\end{equation}

Given that
\[
(\dot{\vec{R}}\cdot\nabla_{\vec{R}})\vec{H}=%
\nabla_{\vec{R}}(\dot{\vec{R}}\cdot\vec{H})-%
 \dot{\vec{R}}\times(\nabla\times\vec{H})
\]
and
\[
\nabla\times\vec{H}=\vec{0}
\]
for any external field, equation (\ref{ecuacion del movimiento
stern-gerlach}) can be written as:
\begin{equation}\label{ecuacion del movimiento stern-gerlach intermedia}
  M\ddot{\vec{R}}=\frac{e}{c}\vec{H}\times\dot{\vec{r}}+%
\frac{e}{c}\nabla_{\vec{R}}(\dot{\vec{R}}\cdot\vec{H})\times\vec{r}-\frac{e}{2\mu
c}\nabla_{\vec{R}} [\vec{H}\cdot(K_L\vec{L}+2\vec{S})],
\end{equation}
or, based on similar reasons:
\begin{equation}\label{ecuacion del movimiento stern-gerlach final}
  M\ddot{\vec{R}}=\frac{e}{c}\vec{H}\times\dot{\vec{r}}+%
\frac{e}{c}\nabla_{\vec{R}}(\dot{\vec{R}}\cdot\vec{H})\times\vec{r}-\frac{e}{2\mu
c}((K_L\vec{L}+2\vec{S})\cdot\nabla_{\vec{R}})\vec{H},
\end{equation}
that can be simplified to
\begin{equation}\label{ecuacion del movimiento stern-gerlach simplificada}
  M\ddot{\vec{R}}=\frac{e}{c}\vec{H}\times\dot{\vec{r}}%
-\frac{e}{2\mu
c}((K_L\vec{L}+2\vec{S})\cdot\nabla_{\vec{R}})\vec{H}
\end{equation}
wherever the component of $\dot{\vec{R}}$ along $\vec{H}$ could be
neglected.

Equation (\ref{ecuacion del movimiento stern-gerlach
simplificada}) shows that the trajectory of the the center of mass
can be deflected by the Stern-Gerlach apparatus even if
$\vec{L}=0$. The term $\frac{e}{c}\vec{H}\times\dot{\vec{r}}$ that
we have encountered before, predicts a spreading of a beam of
atoms in the direction perpendicular to the magnetic field and to
the overall direction of motion, even in a uniform magnetic field.

The gyromagnetic ratio of the internal angular momentum becomes
zero where $m_p=m_e$, as happens with positronium, for which no
contribution to the magnetic momentum results from the internal
angular momentum. In those cases equation (\ref{ecuacion del
movimiento stern-gerlach final}) is simplified to:
\begin{equation}\label{ecuacion del movimiento stern-gerlach positronio}
  M\ddot{\vec{R}}=\frac{e}{c}\vec{H}\times\dot{\vec{r}}+%
\frac{e}{c}\nabla_{\vec{R}}(\dot{\vec{R}}\cdot\vec{H})\times\vec{r}-\frac{e}{\mu
c}(\vec{S}\cdot\nabla_{\vec{R}})\vec{H},
\end{equation}

\section{Atom in an Inhomogeneous Magnetic Field}
We'll consider now a situation where the magnetic field is not
uniform, but remains almost constant inside the atom, in such way
that the vector potential can be smoothly approximated by a linear
function.

The classical Lagrange's Function is:
\begin{equation}\label{nuevo lagrangiano uno}
  L(\vec{r}_p,\vec{r}_e,\vec{v}_p,\vec{v}_e)=%
  \frac{1}{2}m_p v_p^2+\frac{1}{2}m_e v_e^2+%
  \frac{e^2}{\|\vec{r}_p-\vec{r}_e\|}%
  +\frac{e}{c}(\vec{A}(\vec{r}_p)\cdot\vec{v}_p-\vec{A}(\vec{r}_e)\cdot\vec{v}_e)
\end{equation}

We introduce the substitutions (\ref{introduccion del centro de
masas}) and the notation:
\begin{equation}\label{definicion de las deltas}
  \delta\vec{r}_p=-\frac{m_e}{M}\vec{r};\ \ \
  \delta\vec{r}_e=\frac{m_p}{M}\vec{r}.
\end{equation}

To transform (\ref{nuevo lagrangiano uno}) to the system of the
center of mass we'll make use of the relation:
\begin{equation}\label{identidad}
  \vec{v}\cdot\vec{A}(\vec{R}+\delta\vec{R})\approx
  \vec{v}\cdot\vec{A}(\vec{R})+\delta\vec{R}\cdot%
  [(\vec{v}\cdot\nabla_{\vec{R}})\vec{A}(\vec{R})+\vec{v}\times\vec{H}(\vec{R})],
\end{equation}
---valid for any constant vector $\vec{v}$---.

First, we have:

\begin{equation}\label{primer producto}
  \vec{v}_p\cdot\vec{A}(\vec{r}_p)\approx \left(\dot{\vec{R}}-\frac{m_e}{M}\dot{\vec{r}}\right)\cdot\vec{A}(\vec{R})%
  -\frac{m_e}{M}\vec{r}\cdot[(\dot{\vec{R}}\cdot\nabla_{\vec{R}})\vec{A}(\vec{R})]%
\end{equation}
\[
-\frac{m_e}{M}\vec{H}(\vec{R})\cdot(\vec{r}\times\dot{\vec{R}})%
  +\frac{m_e^2}{M^2}\vec{r}\cdot[(\dot{\vec{r}}\cdot\nabla_{\vec{R}})\vec{A}(\vec{R})]%
  +\frac{m_e^2}{M^2}\vec{r}\cdot(\dot{\vec{r}}\times\vec{H}(\vec{R}))
\]
and
\begin{equation}\label{segundo producto}
  \vec{v}_e\cdot\vec{A}(\vec{r}_e)\approx \left(\dot{\vec{R}}+\frac{m_p}{M}\dot{\vec{r}}\right)\cdot\vec{A}(\vec{R})%
  +\frac{m_p}{M}\vec{r}\cdot[(\dot{\vec{R}}\cdot\nabla_{\vec{R}})\vec{A}(\vec{R})],%
\end{equation}
\[
+\frac{m_p}{M}\vec{H}(\vec{R})\cdot(\vec{r}\times\dot{\vec{R}})%
  +\frac{m_p^2}{M^2}\vec{r}\cdot[(\dot{\vec{r}}\cdot\nabla_{\vec{R}})\vec{A}(\vec{R})]%
  +\frac{m_p^2}{M^2}\vec{r}\cdot(\dot{\vec{r}}\times\vec{H}(\vec{R})),
\]
and, therefore:
\begin{equation}\label{diferencia}
  \vec{v}_p\cdot \vec{A}(\vec{r}_p)-\vec{v}_e\cdot
  \vec{A}(\vec{r}_e)\approx -\dot{\vec{r}}\cdot\vec{A}(\vec{R})-%
  \vec{r}\cdot[(\dot{\vec{R}}\cdot\nabla_{\vec{R}})\vec{A}(\vec{R})]
\end{equation}
\[
-\vec{H}(\vec{R})(\vec{R})\cdot(\vec{r}\times\dot{\vec{R}})%
-\frac{m_p-m_e}{M}\vec{r}\cdot((\dot{\vec{r}}\cdot\nabla_{\vec{R}})\vec{A}(\vec{R}))
-\frac{m_p-m_e}{M}\vec{H}(\vec{R})\cdot(\vec{r}\times\dot{\vec{r}})
\]

Now we can rewrite (\ref{nuevo lagrangiano uno}) as:
\begin{equation}\label{nuevo lagrangiano dos}
L(\vec{r}_p,\vec{r}_e,\vec{v}_p,\vec{v}_e)=%
  \frac{1}{2}M \dot{\vec{R}}^2+\frac{1}{2}\mu \dot{\vec{r}}^2+%
  \frac{e^2}{\|\vec{r}_p-\vec{r}_e\|}%
\end{equation}
\[
  +\frac{e}{c}[ -\dot{\vec{r}}\cdot\vec{A}(\vec{R})-%
  \vec{r}\cdot[(\dot{\vec{R}}\cdot\nabla_{\vec{R}})\vec{A}(\vec{R})]
  -\vec{H}(\vec{R})\cdot(\vec{r}\times\dot{\vec{R}})
\]
\[
-\frac{m_p-m_e}{M}\vec{r}\cdot((\dot{\vec{r}}\cdot\nabla_{\vec{R}})\vec{A}(\vec{R}))
-\frac{m_p-m_e}{M}\vec{H}(\vec{R})\cdot(\vec{r}\times\dot{\vec{r}})]
\]
Given that
\begin{equation}\label{simplificacion}
-\dot{\vec{r}}\cdot\vec{A}(\vec{R})-%
  \vec{r}\cdot[(\dot{\vec{R}}\cdot\nabla_{\vec{R}})\vec{A}(\vec{R})]=-%
  \frac{d(\vec{r}\cdot\vec{A}(\vec{R}))}{dt}
\end{equation}
(\ref{nuevo lagrangiano dos}) can be simplified to:
\begin{equation}\label{nuevo lagrangiano tres}
L(\vec{r}_p,\vec{r}_e,\vec{v}_p,\vec{v}_e)=%
    \frac{1}{2}M \dot{\vec{R}}^2+\frac{1}{2}\mu \dot{\vec{r}}^2+%
  \frac{e^2}{\|\vec{r}_p-\vec{r}_e\|}%
\end{equation}
\[
  +\frac{e}{c}[
  -\vec{H}(\vec{R})\cdot(\vec{r}\times\dot{\vec{R}})%
  -\frac{m_p-m_e}{M}\vec{r}\cdot((\dot{\vec{r}}\cdot\nabla_{\vec{R}}))\vec{A}(\vec{R})
\]
\[
-\frac{m_p-m_e}{M}\vec{H}(\vec{R})\cdot(\vec{r}\times\dot{\vec{r}})]
\]

Let's use tensors for the analysis of the term
$\vec{r}\cdot((\dot{\vec{r}}\cdot\nabla_{\vec{R}})\vec{A}(\vec{R}))$:
\[
\vec{r}\cdot((\dot{\vec{r}}\cdot\nabla_{\vec{R}})\vec{A}(\vec{R}))=%
r_i\dot{r}_j\partial_j
A_i=\frac{r_i\dot{r}_j-r_j\dot{r}_i}{2}\partial_j A_i+
\frac{r_i\dot{r}_j+r_j\dot{r}_i}{2}\partial_j A_i=
\]
\[
\frac{\epsilon_{ijk}\epsilon_{kab}}{2}r_a\dot{r}_b\partial_j
A_i+\frac{r_i\dot{r}_j+r_j\dot{r}_i}{2}\partial_j A_i=
\]
\[
-\frac{1}{2}\vec{H}\cdot(\vec{r}\times\dot{\vec{r}})+%
\frac{\vec{r}\cdot(\dot{\vec{r}}\cdot\nabla_{\vec{R}})+
\dot{\vec{r}}\cdot(\vec{r}\cdot\nabla_{\vec{R}})
}{2}\vec{A}(\vec{R})
\]

The Lagrange's Function takes the form:
\begin{equation}\label{nuevo lagrangiano cuatro}
L(\vec{r}_p,\vec{r}_e,\vec{v}_p,\vec{v}_e)=%
    \frac{1}{2}M \dot{\vec{R}}^2+\frac{1}{2}\mu \dot{\vec{r}}^2+%
  \frac{e^2}{\|\vec{r}_p-\vec{r}_e\|}%
\end{equation}
\[
  +\frac{e}{c}[
  -\vec{H}(\vec{R})\cdot(\vec{r}\times\dot{\vec{R}})%
  -\frac{m_p-m_e}{2M}\vec{H}(\vec{R})\cdot(\vec{r}\times\dot{\vec{r}})
\]
\[
-\frac{m_p-m_e}{M}\frac{\vec{r}\cdot(\dot{\vec{r}}\cdot\nabla_{\vec{R}})+
\dot{\vec{r}}\cdot(\vec{r}\cdot\nabla_{\vec{R}})
}{2}\vec{A}(\vec{R})]
\]
Further, we write:
\[
\frac{\vec{r}\cdot(\dot{\vec{r}}\cdot\nabla_{\vec{R}})+
\dot{\vec{r}}\cdot(\vec{r}\cdot\nabla_{\vec{R}})
}{2}\vec{A}(\vec{R})=
\]
\[
\frac{1}{2}\frac{d(\vec{r}\cdot(\vec{r}\cdot\nabla_{\vec{R}})\vec{A}(\vec{R}))}{dt}%
-(\vec{r}\cdot(\vec{r}\cdot\nabla_{\vec{R}}))(\dot{\vec{R}}\cdot\nabla_{\vec{R}})\vec{A}(\vec{R})
\]
demonstrating that the last term of the Lagrange's Function is
equal to a total derivative plus a term of the second order---in
the atomic dimensions---that can be neglected since we have
supposed that the vector potential can be smoothly approximated by
a linear function inside the atom. Therefore:
\begin{equation}\label{nuevo lagrangiano seis}
L(\vec{r}_p,\vec{r}_e,\vec{v}_p,\vec{v}_e)=%
    \frac{1}{2}M \dot{\vec{R}}^2+\frac{1}{2}\mu \dot{\vec{r}}^2+%
  \frac{e^2}{\|\vec{r}_p-\vec{r}_e\|}%
\end{equation}
\[
  -\frac{e}{c}[
  \vec{H}(\vec{R})\cdot(\vec{r}\times\dot{\vec{R}})%
  +\frac{m_p-m_e}{2M}\vec{H}(\vec{R})\cdot(\vec{r}\times\dot{\vec{r}})],
\]
which is the same as (\ref{cuarta lagrangiana}), but with a
magnetic field that depends on the coordinates of the center of
mass, confirming our claim that classical mechanics---and
therefore Schr\"odinger theory also, as follows from Ehrenfest's
Theorem---predicts the result of the Stern-Gerlach experiment.

The Hamilton's Function and the Hamiltonian Operator take the same
form as (\ref{hamiltoniana}) and (\ref{hamiltoniano}),
respectively---where $\vec{H}$ is replaced by $\vec{H}(\vec{R})$.

\section{On the Classical Magnetic Field of the Hydrogen Atom}
We'll estimate the magnetic field produced by the classical
hydrogen atom starting from the classical low-speed-short-distance
approximation of the vector potential of a charge $q$ that moves
along the trajectory $\vec{r}(t)$:
\begin{equation}\label{vector potential formula}
  \vec{A}(\vec{x})=\frac{q}{c}\frac{\dot{\vec{r}}}{\|\vec{x}-\vec{r}\|}
\end{equation}
Obviously, the results wont be valid inside the atom or too far
from it.

From the superposition principle:
\begin{equation}\label{potencial vectorial uno}
  \vec{A}(\vec{x})=\vec{A}_p(\vec{x})+\vec{A}_e(\vec{x}),
\end{equation}
where
\begin{equation}\label{potenciales de proton y electron}
  \vec{A}_p(\vec{x})=\frac{e}{c}\frac{\vec{v}_p}{\|\vec{x}-\vec{r}_p\|}\
  \textbf{and} \ \vec{A}_e(\vec{x})=-\frac{e}{c}\frac{\vec{v}_e}{\|\vec{x}-\vec{r}_e\|}
\end{equation}

First we do the substitutions (\ref{introduccion del centro de
masas}). The result is
\begin{equation}\label{potencial del proton uno}
  \vec{A}_p(\vec{x})=%
  \frac{e}{c}
  \frac{\dot{\vec{R}}-\frac{m_e}{M}\dot{\vec{r}}}
  {\|
  \vec{x}-\vec{R}+\frac{m_e}{M}\vec{r}
  \|
  }
\end{equation}
and
\begin{equation}\label{potencial del electron uno}
  \vec{A}_e(\vec{x})=-%
  \frac{e}{c}
  \frac{\dot{\vec{R}}+\frac{m_p}{M}\dot{\vec{r}}}
  {\|
  \vec{x}-\vec{R}-\frac{m_p}{M}\vec{r}
  \|
  }
\end{equation}

We'll suppose that $\|\vec{x}-\vec{R}\|>\|\vec{r}\|$, in %
such way that the following approximations are reliable:
\[
\frac{1}{\|\vec{x}-\vec{R}+\frac{m_e}{M}\vec{r}\|}\approx%
\frac{1}{\|\vec{x}-\vec{R}\|}-\frac{m_e}{M}%
\frac{(\vec{x}-\vec{R})\cdot\vec{r}}{\|\vec{x}-\vec{R}\|^3}
\]
\[
\frac{1}{\|\vec{x}-\vec{R}-\frac{m_p}{M}\vec{r}\|}\approx%
\frac{1}{\|\vec{x}-\vec{R}\|}+\frac{m_p}{M}%
\frac{(\vec{x}-\vec{R})\cdot\vec{r}}{\|\vec{x}-\vec{R}\|^3}
\]

Then we can show that:
\begin{equation}\label{potencial del proton dos}
  \vec{A}_p(\vec{x})=\frac{e}{c}\left(%
  \frac{\dot{\vec{R}}}{\|\vec{x}-\vec{R}\|}
  -\frac{m_e}{M}\frac{\dot{\vec{r}}}{\|\vec{x}-\vec{R}\|}
  -\frac{m_e}{M}\frac{(\vec{x}-\vec{R})\cdot\vec{r}}{\|\vec{x}-\vec{R}\|^3}\dot{\vec{R}}
  +\frac{m_e^2}{M^2}\frac{(\vec{x}-\vec{R})\cdot\vec{r}}{\|\vec{x}-\vec{R}\|^3}\dot{\vec{r}}
  \right)
\end{equation}
\begin{equation}\label{potencial del electron dos}
  \vec{A}_e(\vec{x})=\frac{e}{c}\left(%
  -\frac{\dot{\vec{R}}}{\|\vec{x}-\vec{R}\|}
  -\frac{m_p}{M}\frac{\dot{\vec{r}}}{\|\vec{x}-\vec{R}\|}
  -\frac{m_p}{M}\frac{(\vec{x}-\vec{R})\cdot\vec{r}}{\|\vec{x}-\vec{R}\|^3}\dot{\vec{R}}
  -\frac{m_p^2}{M^2}\frac{(\vec{x}-\vec{R})\cdot\vec{r}}{\|\vec{x}-\vec{R}\|^3}\dot{\vec{r}}
  \right)
\end{equation}
Therefore:
\begin{equation}\label{potencial vectorial dos}
  \vec{A}(\vec{x})=\frac{e}{c}\left(
  -\frac{\dot{\vec{r}}}{\|\vec{x}-\vec{R}\|}
  -\frac{(\vec{x}-\vec{R})\cdot\vec{r}}{\|\vec{x}-\vec{R}\|^3}\dot{\vec{R}}
  -K_L\frac{(\vec{x}-\vec{R})\cdot\vec{r}}{\|\vec{x}-\vec{R}\|^3}\dot{\vec{r}}
  \right)
\end{equation}
where
\[
K_L=\frac{m_p-m_e}{M}.
\]

In the common treatment of this problem, it's assumed that
$\dot{\vec{R}}=\vec{0}$---and, therefore, that the atom is
actually in rest. This allows to omit the second term, and the
first, after time-averaging. The time average of the vector
potential is then taken as:
\begin{equation}\label{potencial vectorial incorrecto}
 \langle \vec{A}(\vec{x})\rangle=-\frac{e}{c}
  K_L\frac{(\vec{x}-\vec{R})\cdot\vec{r}}{\|\vec{x}-\vec{R}\|^3}\dot{\vec{r}}
\end{equation}

Further it is noticed that:
\[
(x_j-R_j)r_j\dot{r}_i=\frac{(x_j-R_j)(r_j\dot{r}_i-r_i\dot{r}_j)}{2}+%
\frac{(x_j-R_j)(r_j\dot{r}_i+r_i\dot{r}_j)}{2}%
\]
\[
=\frac{(x_j-R_j)(r_j\dot{r}_i-r_i\dot{r}_j)}{2}+%
\frac{1}{2}(x_j-R_j)\frac{d(r_ir_j)}{dt}
\]

The last term is omitted through time-averaging and so the %
usual relation between angular momentum and magnetic moment
emerges, since:
\begin{equation}\label{common relation}
  \langle\vec{A}(\vec{x})\rangle=-\frac{e}{2\mu c}K_L%
  \frac{\vec{L}\times(\vec{x}-\vec{R})}{\|\vec{x}-\vec{R}\|^3}
\end{equation}
and, therefore:
\begin{equation}\label{common magnetic moment}
  \vec{\mu}=-\frac{e}{2\mu c}K_L\vec{L}.
\end{equation}

It is necessary to stress the fact that the magnetic moment as
defined by eq. (\ref{common magnetic moment}), that proceeds from
an average of dynamical variables, can only be used to estimate,
not to compute, the instantaneous energy associated to the
interaction of the atom with an external magnetic field, and,
therefore, by itself, is not acceptable for quantization.

Furthermore, even at this statistical level, we have a correction
to the gyromagnetic ratio:
\begin{equation}\label{gyromagnetic ratio}
  g=\frac{e}{2\mu c}K_L,
\end{equation}
which is important, because $g=0$ for the positronium atom.

To simplify the task of understanding the magnetic field %
associated to the potential (\ref{potencial vectorial dos}) %
we write it as a sum of two terms:
\begin{equation}\label{analisis del potencial}
  \vec{A}(\vec{x})=\vec{A}_1(\vec{x})+\vec{A}_2(\vec{x})
\end{equation}
where
\begin{equation}\label{primer termino}
  \vec{A}_1(\vec{x})=
  -\frac{e}{c}\frac{\dot{\vec{r}}}{\|\vec{x}-\vec{R}\|}
\end{equation}
and
\begin{equation}\label{segundo termino}
  \vec{A}_2(\vec{x})=
  -\frac{e}{c}\frac{(\vec{x}-\vec{R})\cdot\vec{r}}{\|\vec{x}-\vec{R}\|^3}(\dot{\vec{R}}+K_L\dot{\vec{r}})
\end{equation}

The field associated to the first term:
\begin{equation}\label{aura magnetica}
  \vec{H}_1(\vec{x})=-\frac{e}{c}\frac{\dot{\vec{r}}\times(\vec{x}-\vec{R})}%
  {\|\vec{x}-\vec{R}\|^3},
\end{equation}
is like a magnetic spinning belt, surrounding the atom, with the
axis parallel to $\dot{\vec{r}}$. The intensity of this field
decreases as $\|\vec{x}-\vec{R}\|^{-2}$. Therefore, it has a
longer range than the dipolar terms and cannot be compensated by
them.

The second term (\ref{segundo termino}) produces the field:
\begin{equation}\label{campo dipolar}
\vec{H}_2(\vec{x})=-\frac{e}{c}(\dot{\vec{R}}+K_L\dot{\vec{r}})\times\frac{3(\vec{r}\cdot(\vec{x}-\vec{R}))%
(\vec{x}-\vec{R})-\|\vec{x}-\vec{R}\|^2\vec{r}}%
{\|\vec{x}-\vec{R}\|^5},
\end{equation}
that can be written as:
\begin{equation}\label{campo dipolar magnetico}
\vec{H}_2(\vec{x})=\frac{1}{c}(\dot{\vec{R}}+K_L\dot{\vec{r}})\times
\vec{E}_{\vec{p}}(\vec{x}),
\end{equation}
where
\begin{equation}\label{campo dipolar electrico}
  \vec{E}_{\vec{p}}=\frac{3(\vec{p}\cdot(\vec{x}-\vec{R}))%
(\vec{x}-\vec{R})-\|\vec{x}-\vec{R}\|^2\vec{p}}{\|\vec{x}-\vec{R}\|^5}
\end{equation}
is the electric field associated to the electric dipole
$-e\vec{r}$.

The electric field---under the same approximations---is given by:
\begin{equation}\label{electric field}
  \vec{E}(\vec{x})=\vec{E}_{\vec{p}}(\vec{x})-\frac{1}{c}\frac{\partial \vec{A}(\vec{x})}{\partial t}
\end{equation}


\begin{thebibliography}{ccc}
\bibitem{OCHAVOYA} O. Chavoya-Aceves; \emph{Remarks on the Theory of Angular
Momentum.}; arXiv:quant-ph/0305049 (2003).
\bibitem{WEYL} H. Weyl; \emph{The Theory of Groups and Quantum
Mechanics}; Dover (1950).
\bibitem{MESSIA} A. Messiah; \emph{Quantum Mechanics}; Dover
(1999).%
\bibitem{GARRAWAY-STENHOLM} B. M. Garraway, S. Stenholm; \emph{Does the flying electron spin?};
Contemp. Phys. \textbf{43}, 147 (2002).
\bibitem{BOHM} D. Bohm;\emph{Quantum Theory}; Dover (1989).
\end{thebibliography}
\end{document}